\documentclass[prb,showpacs,twocolumn]{revtex4}
\usepackage{epsfig}
\usepackage{graphics}
\usepackage{graphicx}
\usepackage{dcolumn}
\usepackage{bm}

\newcommand{\sba}{\begin{subeqnarray}}
\newcommand{\sea}{\end{subeqnarray}}

\def\cm-1{cm$^{-1}$}

\begin{document}

\title{Single-site Anderson Model. I Diagrammatic theory}
\author{V.\ A.\ Moskalenko$^{1,2}$}
\email{moskalen@thsun1.jinr.ru}
\author{P.\ Entel$^{3}$}
\author{D.\ F.\ Digor$^{1}$}\author{L.\ A.\ Dohotaru$^{4}$} \author{R.\ Citro$^{5}$}
\affiliation{$^{1}$Institute of Applied Physics, Moldova Academy
of Sciences, Chisinau 2028, Moldova} \affiliation{$^{2}$BLTP,
Joint Institute for Nuclear Research, 141980 Dubna, Russia}
\affiliation{$^{3}$University of Duisburg-Essen, 47048 Duisburg,
Germany}
\affiliation{$^{4}$Technical University, Chisinau 2004,
Moldova}
\affiliation{$^{4}$Dipartimento di Fisica "E. R. Caianiello", Universit\'{a} degli Studi
di Salerno and CNISM, Unit\'{a} di ricerca di Salerno, Via S. Allende, 84081 Baronissi (SA), Italy}

\date{\today}

\begin{abstract}
%%%%%%%%%%%%%%%%
%
The diagrammatic theory is proposed for the strongly correlated
impurity Anderson model. The strongly correlated impurity
electrons are hybridized with free conduction electrons. For this
system the new diagrammatic approach is formulated. The linked
cluster theorem for vacuum diagrams is proved and the Dyson type
equations for electron propagators of both electron subsystems are
established, together with such equations for mixed propagators.
The approximations based on the summing the infinite series of
diagrams are proposed, which close the system of equations and
permit the investigation of the system's properties.
%%%%%%%%%%%%%%
\end{abstract}

\pacs{78.30.Am, 74.72Dn, 75.30.Gw, 75.50.Ee}

\maketitle

%%%%%%%%%%%%%%%%%%%%%%%%%

\section{Introduction}

%%%%%%%%%%%%%%%%%%%%%%%%%

The study of strongly-correlated electron systems become in the
last decade one of the most active fields of condensed matter
physics. The properties of these systems can not be described by
Fermi liquid theory. One of the important models of strongly
correlated electrons is the single-site or impurity model
introduced by Anderson$^{[1]}$ in the 1961 and discussed
intensively in a lot of papers$^{[2-15]}$. It is a model for a
system of free conduction electrons that interact with the system
of local spin, treated as just another electrons of $d$- or $f$-
shells of an impurity atom. The impurity electrons are strongly
correlated because of strong Coulomb repulsion and they undergo
the exchange and hybridization with conduction electrons. This
model has some properties similar to those of Kondo model having
more interesting physics$^{[16-18]}$. It has the application for
heavy fermion systems where the local impurity orbital is $f$ -
orbital. Investigations of impurity Anderson model have used
intensively the methods and results obtained for Kondo model by
Nagaoka$^{[18]}$ and other authors$^{[19,20]}$. All the cited
papers are based on the method of equation of motions for
retarded and advanced quantum Green's functions proposed by
Bogoliubov and Tiablikov$^{[21]}$ and developed in papers$
^{[22-24]}$.

The first attempt to develop the diagrammatic theory for this
problem was realized in the paper$^{[25]}$. These authors used the
expansion by cumulants for averages of products of Hubbard
transfer operators and their algebra.

With introduction of Dynamical Mean Field Theory the interest for
Anderson impurity model increases because infinite dimensional
lattice models can be mapped onto effective impurity models
together with a self-consistency condition$^{[26,27]}$.

The Hamiltonian of the model is written as
%
%%%%%%%%%%%%%%%%
\begin{eqnarray}
%%%%%%%%%%%%%%%%
%
H &=&H_{0}+H_{int},  \nonumber \\
H_{0} &=&H_{0}^{c}+H_{0}^{f},  \nonumber \\
H_{0}^{c} &=&\sum\limits_{\mathbf{k}\sigma }\epsilon (\mathbf{k})\
C_{
\mathbf{k}\sigma }^{+}C_{\mathbf{k}\sigma }.  \nonumber \\
H_{0}^{f} &=&\epsilon _{f}\sum\limits_{\sigma }f_{\sigma }^{+}f_{\sigma
}+Un_{\uparrow }^{f}n_{\downarrow }^{f}, \\
H_{int} &=&\frac{1}{\sqrt{N}}\sum\limits_{\mathbf{k}\sigma }\left(
V_{ \mathbf{k}\sigma }f_{\sigma }^{+}C_{\mathbf{k}\sigma
}+V_{\mathbf{k}\sigma
}^{\ast }C_{\mathbf{k}\sigma }^{+}f_{\sigma }\right) ,  \nonumber \\
n_{\sigma }^{f} &=&f_{\sigma }^{+}f_{\sigma },  \nonumber
\label{1}
%
%%%%%%%%%%%%%%
\end{eqnarray}
%%%%%%%%%%%%%%
%
where $C_{\mathbf{k}\sigma }(C_{\mathbf{k}\sigma }^{+})$ and
$f_{\sigma }(f_{\sigma }^{+})$ - annihilation (creation) operators
of conduction and impurity electrons with spin $\sigma $\
correspondingly. $\epsilon (\mathbf{k})$ is the kinetic energy of
the conduction band state $(\mathbf{k},\sigma )$ , $\epsilon _{f}\
$is the local energy of $f$ - electrons, $U$ - is the on-site
Coulomb repulsion of the impurity electrons and $N$ is the number
of lattice sites. $H_{int}$ is the hybridization interaction
between conduction and localized electrons. Summation over
$\mathbf{k}$  will be changed to an integral over the energy
$\epsilon (\mathbf{k})$ with the density of state $\rho
_{0}(\epsilon )$ of conduction electrons and the matrix elements
will be considered as the function of energy $V(\epsilon )$.
Because of the hybridization term of the Hamiltonian down and up
spins of conduction electrons come and go in the local orbital and
there is no appearance of spin flip process. Thus the important
parameters of the Anderson model are the band width $W$, the
conduction density of states $\rho (\epsilon ),$ the local site
energy $\epsilon _{f}$ and the on-site Coulomb interaction $U$.
The electron energies are counted of chemical potential $\mu $\ of
the system: $\epsilon (\mathbf{k})=\xi (\mathbf{k})-\mu ,\qquad
\epsilon _{f}= \overline{\epsilon }_{f}-\mu $ . There is also an
energy parameter $\Gamma (\epsilon )$ associated with the
hybridization term
%
%%%%%%%%%%%%%%%%
\begin{equation}
%%%%%%%%%%%%%%%%
%
\Gamma (\epsilon )=\frac{\pi
}{N}\sum\limits_{\mathbf{k}}V_{\mathbf{k} }^{2}\delta (\epsilon
-\epsilon (\mathbf{k}))=\pi V^{2}(\epsilon )\rho_{0}(\epsilon
).\label{2}
%
%%%%%%%%%%%%%%
\end{equation}
%%%%%%%%%%%%%%
%
This function is assumed to be a constant, independent of energy.
The term in the Hamiltonian involving $U$ comes from on-site
Coulomb interaction between two impurity electrons. $U$\ it is far
to large to be treated by perturbation theory. It must be included
in $Ho$ which is non interacting Hamiltonian. The existence of
this term invalidates Wick's theorem for local electrons.
Therefore, first of all, we formulate the generalized Wick's
theorem (GWT) for local electrons, preserving the ordinary Wick
theorem for conduction electrons. Our GWT really is the identity
which determines the irreducible Green's functions or Kubo
cumulants. Such definitions have already been used by us for
discussing the properties of one-band Hubbard model$^{[28-30]}$
and the formulation of the new diagram technique for it$
^{[31-34]}$.

In Section II, we start by introducing the temperature Green's
functions for the conduction and impurity electrons in interaction
representation, formulate the generalized Wick theorem and provide
explicit examples of diagram calculation for thermodynamical
potential and full propagators. The results are analyzed in
Section III and compared to the other data in Section IV. Some
approximations are discussed in Section V and in Section VI there
are the conclusions.

%%%%%%%%%%%%%%%%%%%%%%%%%

\section{Diagrammatical theory}

%%%%%%%%%%%%%%%%%%%%%%%%%

The Matsubara renormalized Green's functions of conduction and impurity
electrons in interaction representation have the form:
%
%%%%%%%%%%%%%%%%
\begin{eqnarray}
%%%%%%%%%%%%%%%%
%
G(\mathbf{k,}\sigma ,\tau \mid \mathbf{k}^{\prime },\sigma
,^{\prime }\tau ^{\prime }) &=&-\ \left\langle
TC_{\mathbf{k}\sigma }(\tau )\overline{C}_{ \mathbf{k}^{\prime
}\sigma ^{\prime }}(\tau ^{\prime })U(\beta
)\right\rangle _{0}^{c},  \nonumber \\
g(\sigma ,\tau \mid \sigma ^{\prime },\tau ^{\prime }) &=&-\
\left\langle Tf_{\sigma }(\tau )\overline{f}_{\sigma ^{\prime
}}(\tau ^{\prime })U(\beta )\right\rangle _{0}^{c}.\label{3}
%
%%%%%%%%%%%%%%
\end{eqnarray}
%%%%%%%%%%%%%%
%

Besides them there are also anomalous ones:
%
%%%%%%%%%%%%%%%%
\begin{eqnarray}
%%%%%%%%%%%%%%%%
%
F(\mathbf{k,}\sigma ,\tau \mid -\mathbf{k,}-\sigma ^{\prime },\tau
^{\prime }) &=&-\ \left\langle TC_{\mathbf{k}\sigma }(\tau
)C_{-\mathbf{k}^{\prime }-\sigma ^{\prime }}(\tau ^{\prime
})U(\beta )\right\rangle _{0}^{c},
\nonumber \\
\overline{F}(-\mathbf{k,}-\sigma ,\tau \mid \mathbf{k}^{\prime },\sigma
^{\prime },\tau ^{\prime }) &=&-\ \left\langle T\overline{C}_{-\mathbf{k}
-\sigma }(\tau )\overline{C}_{\mathbf{k}^{\prime }\sigma ^{\prime }}(\tau
^{\prime })U(\beta )\right\rangle _{0}^{c},  \nonumber \\
f(\sigma ,\tau \mid -\sigma ^{\prime },\tau ^{\prime }) &=& -\ \left\langle
Tf_{\sigma }(\tau )f_{-\sigma ^{\prime }}(\tau ^{\prime })U(\beta
)\right\rangle _{0}^{c},  \nonumber \\
\overline{f}(-\sigma ,\tau \mid \sigma ^{\prime },\tau ^{\prime })
&=&-\ \left\langle T\overline{f}_{-\sigma }(\tau
)\overline{f}_{\sigma ^{\prime }}(\tau ^{\prime })U(\beta
)\right\rangle _{0}^{c},  \nonumber \\\label{4}
%
%%%%%%%%%%%%%%
\end{eqnarray}
%%%%%%%%%%%%%%
%
if the system is in superconducting state. Here $\tau $\ and $\tau
^{\prime } $ stand for imaginary time with $0<\tau <\beta $,
$\beta $ - inverse temperature and T is the chronological ordering
operator. The evolution operator $U(\beta )$ is given by
%
%%%%%%%%%%%%%%%%
\begin{widetext}
%%%%%%%%%%%%%%%%
%
%%%%%%%%%%%%%%%%
\begin{eqnarray}
%%%%%%%%%%%%%%%%
%
U(\beta ) & = & T\exp (-\int\limits_{0}^{\beta }H_{int}(\tau
)d\tau )=\sum\limits_{n=0}^{\infty
}\frac{(-1)^{n}}{n!}\int\limits_{0}^{\beta }d\tau
_{1}...\int\limits_{0}^{\beta }d\tau _{n}T(H_{int}(\tau
_{1})...H_{int}(\tau _{n})).\label{5}
%
%%%%%%%%%%%%%%
\end{eqnarray}
%%%%%%%%%%%%%%
%
%%%%%%%%%%%%%%
\end{widetext}
%%%%%%%%%%%%%%
%
The statistical averaging is carried out in (3) and (4) with respect to the
zero-order density matrix of the conduction and impurity electrons.
%
%%%%%%%%%%%%%%%%
\begin{equation}
%%%%%%%%%%%%%%%%
%
\frac{e^{-\beta H_{0}}}{Tre^{-\beta H_{0}}}=\frac{e^{-\beta
H_{0}^{c}}}{ Tre^{-\beta H_{0}^{c}}}\times\frac{e^{-\beta
H_{0}^{f}}}{Tre^{-\beta H_{0}^{f}}}.\label{6}
%
%%%%%%%%%%%%%%
\end{equation}
%%%%%%%%%%%%%%
%

The thermodynamic perturbation theory for $H_{int}$ requires the
generalization adequate for calculation of the statistical
averages of the $ T\ $- products of localized $f$ - electron
operators. This necessity appears for the reason that cannot be
diagonalized with free $f$ - electron operators. This Hamiltonian
can be diagonalized by using the algebra of Hubbard$^{[28-30]}$
transfer operators $\chi ^{mn}=\left\vert m><n\right\vert $ when
the $\mid $\/$m$\/$>$ with $m=-1,0,1,2$ enumerates four states of
the impurity atom: $\mid $\/$0>$ - is the empty or vacuum state
with energy $E_{0}=0$, the $\mid $\/$1>$ and $\mid $\/$-1>$ or
$\mid $ \/$\uparrow >$ and $\mid $\/$\downarrow >$ are the states
with one particle with energy $E_{\sigma }=\epsilon _{f}$ and spin
$\sigma =\pm 1$ and the state $\mid $\/$2>=\mid $\/$\uparrow
\downarrow >$ contains two $f$ - electrons with opposite spins and
the energy $E_{2}=U+2\epsilon _{f}$. By using the relation
%
%%%%%%%%%%%%%%%%
\begin{equation}
%%%%%%%%%%%%%%%%
%
f_{\sigma }=\chi ^{0\sigma }+\sigma \chi ^{\overline{\sigma
}2},\label{7}
%
%%%%%%%%%%%%%%
\end{equation}
%%%%%%%%%%%%%%
%
we obtain the diagonalized form of the impurity Hamiltonian
%
%%%%%%%%%%%%%%%%
\begin{equation}
%%%%%%%%%%%%%%%%
%
H_{0}^{f}=\sum\limits_{n=-1}^{2}E_{n}\chi ^{nn},\qquad
\sum\limits_{n=-1}^{2}\chi ^{nn}=1.\label{8}
%
%%%%%%%%%%%%%%
\end{equation}
%%%%%%%%%%%%%%
%

In zero order approximation, when we neglect the process of
hybridization of the conduction and impurity electrons, the
corresponding Green's functions have the form $(\omega \equiv
\omega_{n} =(2n+1)\pi /\beta )$
%
%%%%%%%%%%%%%%%%
\begin{eqnarray}
%%%%%%%%%%%%%%%%
%
G_{\sigma \sigma ^{\prime }}^{0}(\mathbf{k},\mathbf{k}^{\prime
}\mid i\omega ) &=&\delta _{\sigma \sigma ^{\prime }}\delta
_{\mathbf{kk}^{\prime }}\frac{
1 }{i\omega -\epsilon (\mathbf{k)}},  \nonumber \\
g_{\sigma \sigma^{\prime}}^0=\delta_{\sigma \sigma^{\prime}}g_{\sigma }^{0}(i\omega ) &=&\frac{1-n_{\overline{\sigma
}}}{\lambda _{\sigma }(i\omega )}+\frac{n_{\overline{\sigma
}}}{\overline{\lambda }_{ \overline{\sigma }}(i\omega )},\label{9}
%
%%%%%%%%%%%%%%
\end{eqnarray}
%%%%%%%%%%%%%%
%
where $(\overline{\sigma }=-\sigma )$
%
%%%%%%%%%%%%%%%%
\begin{eqnarray}
%%%%%%%%%%%%%%%%
%
\lambda _{\sigma }(i\omega ) &=&i\omega +E_{0}-E_{\sigma }  \nonumber \\
\overline{\lambda} _{\overline{\sigma }}(i\omega ) &=&i\omega +E_{\overline{\sigma }
}-E_{2},  \nonumber \\
Z_{0} &=&e^{-\beta E_{0}}+e^{-\beta E_{\sigma }}+e^{-\beta E_{\overline{
\sigma }}}+e^{-\beta E_{2}},  \nonumber \\
n_{\overline{\sigma }} &=&\frac{e^{-\beta E_{\overline{\sigma }}}+e^{-\beta
E_{2}}}{Z_{0}},  \nonumber \\
1-n_{\overline{\sigma }} &=&\frac{e^{-\beta E_{0}}+e^{-\beta E_{\sigma }}}{
Z_{0}}.  \nonumber
%
%%%%%%%%%%%%%%
\end{eqnarray}
%%%%%%%%%%%%%%
%

In the case of $f$ - electrons we formulate the identity which is just our
GWT in this simple case:
%
%%%%%%%%%%%%%%%%
\begin{widetext}
%%%%%%%%%%%%%%%%
%
%%%%%%%%%%%%%%%%
\begin{eqnarray}
%%%%%%%%%%%%%%%%
%
\left\langle
Tf_{1}f_{2}\overline{f}_{3}\overline{f}_{4}\right\rangle _{0}& = &
\left\langle Tf_{1}\overline{f}_{4}\right\rangle _{0}\left\langle
Tf_{2}
\overline{f}_{3}\right\rangle _{0} -\left\langle Tf_{1}\overline{f}%
_{3}\right\rangle _{0}\left\langle
Tf_{2}\overline{f}_{4}\right\rangle _{0}+\left\langle
Tf_{1}f_{2}\overline{f}_{3}\overline{f}_{4}\right\rangle
_{0}^{ir},\label{10}
%
%%%%%%%%%%%%%%
\end{eqnarray}
%%%%%%%%%%%%%%
%
or
%
%%%%%%%%%%%%%%%%
\begin{eqnarray}
%%%%%%%%%%%%%%%%
%
g_{2}^{0}(1,2|3,4)&=&
g^{0}(1|4)g^{0}(2|3)-g^{0}(1|3)g^{0}(2|4)+g_{2}^{(0)ir}(1,2|3,4),\label{11}
%
%%%%%%%%%%%%%%
\end{eqnarray}
%%%%%%%%%%%%%%
%
%%%%%%%%%%%%%%
\end{widetext}
%%%%%%%%%%%%%%
%
where $n$ stands for $(\sigma _{n},\tau _{n})$. The generalization
for more complicate averages of type \newline
$g_{n}^{0}(1,...,n\mid n+1,...,2n)=(-1)^{n}\left\langle
Tf_{1}...f_{n} \overline{f}_{n+1}...\overline{f}_{2n}\right\rangle
_{0}$ is straightforward, namely the right - hand part of this
quantity will contain $ n!$ term of ordinary Wick type (chain
diagrams) and also the different products of irreducible functions
with the same total number of operators. The full irreducible
functions in \\
$g_{n}^{0}(1,...,n|n+1,...,2n)$ also appears. For example
$g_{3}^{0}(123|456)$ contains the contribution of $ 3!=6$
terms of ordinary Wick kind, then appear 9 terms of the form $
g^{0}(1|4)g_{2}^{(0)ir}(23|56)$ and the last term is
$g_{3}^{(0)ir}(123|456)$ . The total number of terms is 16. In the
case of $g_{4}^{0}(1234|5678)$ there are $4!=24$ terms of ordinary
Wick kind, the 72 terms of the type $
g^{0}(1|5)g^{0}(2|6)g_{2}^{(0)ir}(34|78)$, then 18 terms of type $
g_{2}^{(0)ir}(12|56)g_{2}^{(0)ir}(34|78)$, then 16 terms of the
form $ g^{0}(1|5)g_{3}^{(0)ir}(234|678)$ and finally one form $
g_{4}^{(0)ir}(1234|5678)$. The total number of terms is 131. The
signs of all these contributions can be easily determined. Thus
the definition of the irreducible Green's functions or Kubo
cumulants is just our GWT. In the absence of Coulomb repulsion $U\
$all these irreducible functions are equal to zero. When $U\neq 0$
they contain all the spin, charge and pairing fluctuations
produced by the strong correlations. These definitions are the
simplification of ones for Hubbard and other lattice models. The
calculation of the simplest irreducible functions for example
$g_{2}^{(0)ir}(12|34)$ is rather cumbersome but straightforward.
It is necessary to find the values of chronological averages for
$4!=24$ different orders of $\tau _{1},\tau _{2},\tau _{3}$ and
$\tau _{4}$ times and then to determine its Fourier representation
%
%%%%%%%%%%%%%%%%
\begin{widetext}
%%%%%%%%%%%%%%%%
%
%%%%%%%%%%%%%%%%
\begin{eqnarray}
%%%%%%%%%%%%%%%%
%
g_{2}^{(0)ir}[\sigma _{1},\tau _{1};\sigma _{2},\tau _{2}|\sigma
_{3},\tau _{3};\sigma _{4},\tau _{4}] & = & \frac{1}{\beta
^{4}}\sum\limits_{\omega _{1}\omega _{2}\omega _{3}\omega
_{4}}\exp (-i\omega _{1}\tau _{1}-i\omega _{2}\tau _{2}+i\omega
_{3}\tau _{3}+i\omega _{4}\tau _{4})\times
\nonumber \\
&& \\
&\times &g_{2}^{(0)ir}[\sigma _{1},i\omega _{1};\sigma
_{2},i\omega _{2}|\sigma _{3},i\omega _{3};\sigma _{4},i\omega
_{4}]. \nonumber \label{12}
%
%%%%%%%%%%%%%%
\end{eqnarray}
%%%%%%%%%%%%%%
%
The Fourier representation conserves the frequencies
%
%%%%%%%%%%%%%%%%
\begin{eqnarray}
%%%%%%%%%%%%%%%%
%
g_{2}^{(0)ir}[\sigma _{1},i\omega _{1};\sigma _{2},i\omega
_{2}|\sigma _{3},i\omega _{3};\sigma _{4},i\omega _{4}] & = &
\beta \delta (\omega
_{1}+\omega _{2}-\omega _{3}-\omega _{4})\times  \nonumber \\
&& \\
&\times &\widetilde{g}_{2}^{(0)ir}[\sigma _{1},i\omega _{1};\sigma
_{2},i\omega _{2}|\sigma _{3},i\omega _{3};\sigma _{4},i\omega
_{1}+i\omega _{2}-i\omega _{3}].  \nonumber \label{13}
%
%%%%%%%%%%%%%%
\end{eqnarray}
%%%%%%%%%%%%%%
%
%%%%%%%%%%%%%%
\end{widetext}
%%%%%%%%%%%%%%
%
There is also the spin conservation $\sigma _{1}+\sigma _{2}=\sigma
_{3}+\sigma _{4}$. Thus we have the rules to deal with chronological
averages of thermodynamic perturbation theory.

%%%%%%%%%%%%%%%%%%%%%%%%%

\section{Thermodynamic potential}

%%%%%%%%%%%%%%%%%%%%%%%%%

First of all we can determine the thermodynamic potential $F$ of
the system
%
%%%%%%%%%%%%%%%%
\begin{eqnarray}
%%%%%%%%%%%%%%%%
%
F &=&F_{0}-\frac{1}{\beta }\ln \left\langle U(\beta )\right\rangle
_{0},
\nonumber \\
&& \\
F_{0} &=&-\frac{1}{\beta }\ln Z_{0}-\frac{2}{\beta
}\sum\limits_{\mathbf{k} }\ln \left[ 1+\exp (-\beta \epsilon
(\mathbf{k)})\right],  \nonumber \label{14}
%
%%%%%%%%%%%%%%%%
\end{eqnarray}
%%%%%%%%%%%%%%%%
%
where $Zo$ refers to the free impurity atom. The diagrams which
determine the thermodynamic potential have not the external lines
and are named vacuum.
%
%%%%%%%%%%%%%%%%%%
\begin{figure*}[t]
%%%%%%%%%%%%%%%%%%
%
\centering
\includegraphics[width=0.75\textwidth,clip]{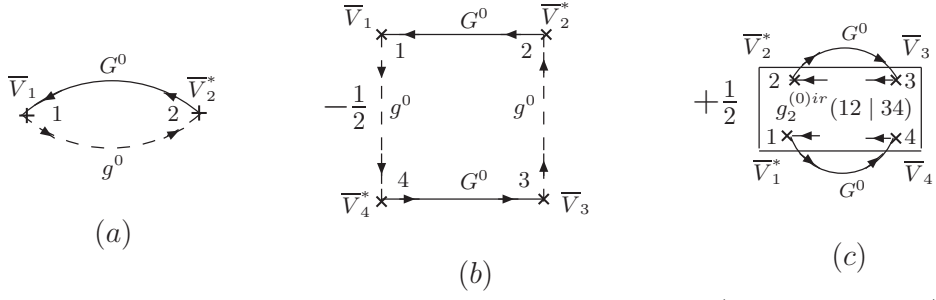}
\vspace{-5mm}
%
%%%%%%%%%%%%%
\caption{ The simplest connected vacuum diagrams in normal state.
The diagram $ a)$ is of second and $b)$, $c)$ of fourth order of
the theory. Here $ \overline{V}_{n}=V_{n}/\sqrt{N}$.
}\label{fig-1} \vspace{-5mm}
\end{figure*}
%%%%%%%%%%%%%
%
%%%%%%%%%%%%%%%%%%
\begin{figure*}[t]
%%%%%%%%%%%%%%%%%%
%
\centering
\includegraphics[width=0.65\textwidth,clip]{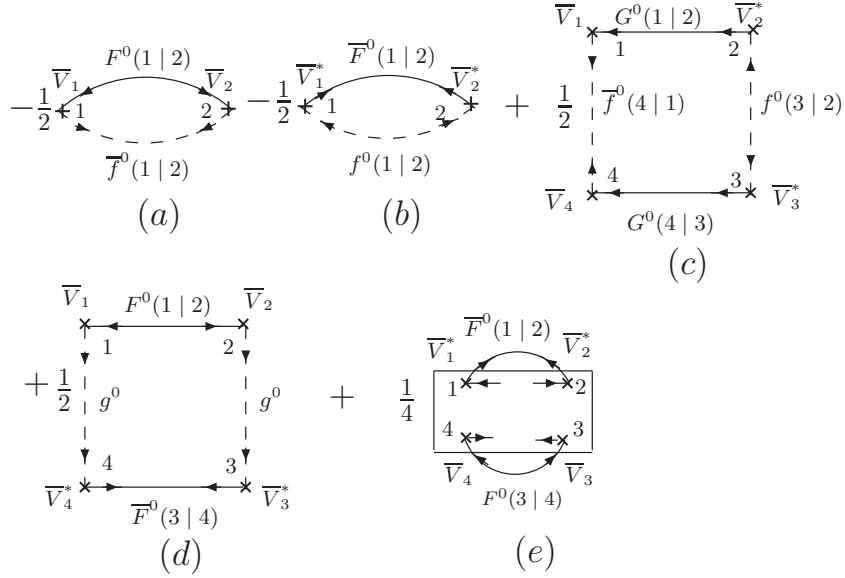}
\caption{ The simplest vacuum anomalous diagrams. The diagrams
$a)$ and $b)$ are of second and $c)$, $d)$ and $e)$ of fourth
order of perturbation theory. }\label{fig-2}
%
%%%%%%%%%%%%%
\end{figure*}
%%%%%%%%%%%%%
%

In $Fig.1$ are shown the simplest vacuum connected diagrams of the
normal state. In the diagrams we shall depict the process of
hybridization of $C$\ and $f$\ electrons. The zero order
propagators of conduction and impurity electrons are represented
by their solid and dashed lines correspondingly. These lines
connect the crosses which depict the impurity states. To crosses
are attached two arrows, one of which is ingoing and other
outgoing. They depict the annihilation and creation electrons
correspondingly. The index $n$ \ means $(\sigma _{n},\tau _{n})$
for impurity and $(\mathbf{k}_{n},\sigma _{n},\tau _{n})$ for
conduction electrons. The rectangles with $2n$\ crosses depict the
irreducible $g_{n}^{(0)ir}$ Green's functions.

Besides the vacuum diagrams of fourth order shown on the $Fig.1$ \
\ $b)$\ and $c)$ there is also one disconnected diagram composed
from two diagrams of the type $Fig.1$ \ $a)$\ \ and containing
additional factor \ $1/2!$. Such situation is repeated in high
order of perturbation theory and permit us to formulate linked
cluster theorem. It has the form
%
%%%%%%%%%%%%%%%%
\begin{equation}
%%%%%%%%%%%%%%%%
%
\left\langle U(\beta )\right\rangle _{0}=\exp \left\langle U(\beta
)\right\rangle _{0}^{c},\label{15}
%
%%%%%%%%%%%%%%
\end{equation}
%%%%%%%%%%%%%%
%
where $\left\langle U(\beta )\right\rangle _{0}^{c}$ contains only connected
diagrams and is equal to zero when hybridization is absent. If we admit the
existence of the pairing mechanism of conduction electrons, thanks the
hybridization, the paring mechanism appear also for impurity electrons. This
mechanism results in appearance of the anomalous propagators of both kind of
electrons.

$Fig.2$ shows some of the simplest connected anomalous vacuum
diagrams. The anomalous propagators are depicted by the thin
(solid and dashed) lines with two opposite directions at the end
of them.
%
%%%%%%%%%%%%%%%%%%
\begin{figure*}[t]
%%%%%%%%%%%%%%%%%%
%
\centering
\includegraphics[width=0.95\textwidth,clip]{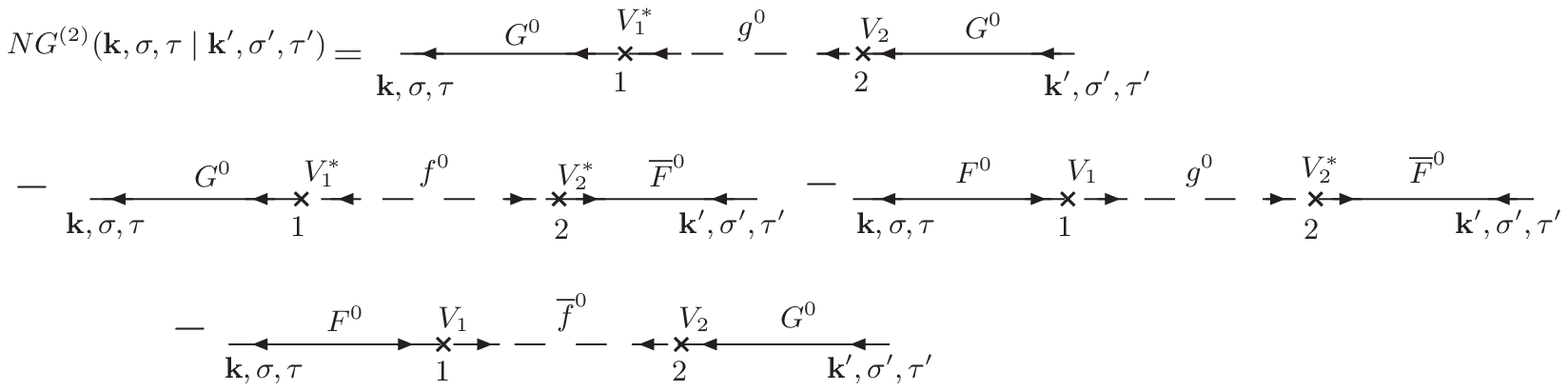}
\caption{ Second order perturbation theory contribution for
conduction electron normal propagator. }\label{fig-3}
%
%%%%%%%%%%%%%
\end{figure*}
%%%%%%%%%%%%%
%
%%%%%%%%%%%%%%%%%%
\begin{figure*}[t]
%%%%%%%%%%%%%%%%%%
%
\centering
\includegraphics[width=0.95\textwidth,clip]{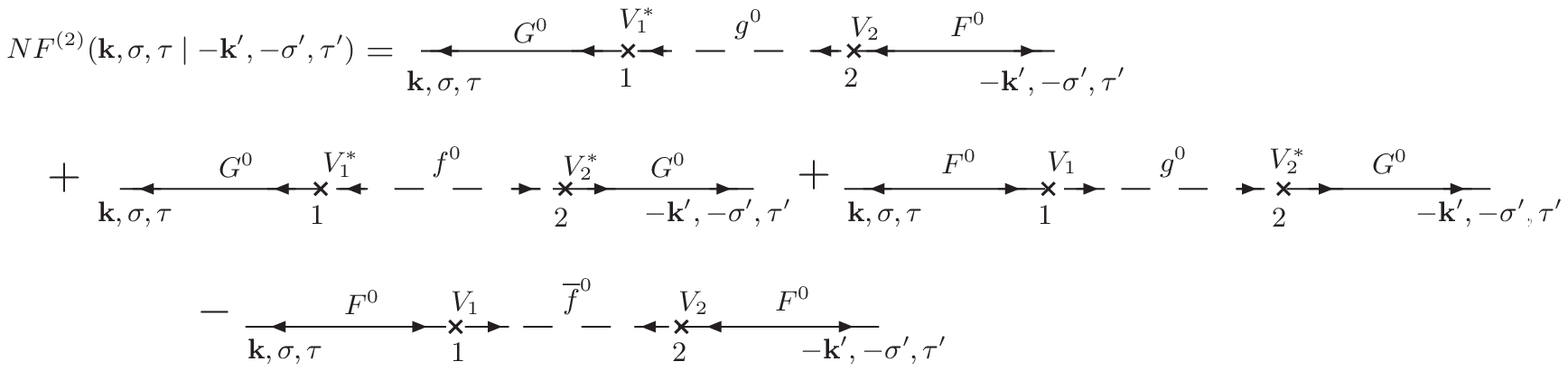}
\caption{ Second order perturbation theory contribution for
conduction electron anomalous propagator. }\label{fig-4}
%
%%%%%%%%%%%%%
\end{figure*}
%%%%%%%%%%%%%
%
%%%%%%%%%%%%%

\section{Renormalized propagators}

%%%%%%%%%%%%%%%%%%%%%%%%%

Now we shall consider the diagrammatical analysis of the
perturbation series for renormalized propagators (3) and (4). The
simplest contributions to such series are represented on the
$Figures\ 3-6$. All such diagrams contain two external points with
attached arrows determined by the arguments of Green's functions
and their kind.On the inner points of diagrams is supposed
summation on $\sigma _{n},\mathbf{k}_{n}$, and integration on
$\tau _{n}$.

In the same second order approximation of perturbation theory the diagrams
for impurity electron propagators contain new diagrammatical elements namely
the irreducible two particle Green's functions. These functions also can be
normal or anomalous. The process of their renormalization will be not
considered by us, supposing the necessity of renormalization only for the
propagators.

In $Fig.5$ the diagrams for impurity electron normal propagator
are shown.

The last two irreducible Green's functions of $ Fig.5$ are
anomalous ones because they contain non equal number of
annihilation and creation $f$-operators enumerated in the left and
right parts about the vertical bare correspondingly. Thanks the
summation of infinite series diagrams the renormalized normal and
anomalous propagators appear and now it is necessary to put equal
to zero the source of electron pairs and simultaneously the bare
$f^{0}$ and $\overline{f}^{0}$\ together with anomalous
irreducible Green's functions. The corresponding contribution to
the anomalous impurity electron function $f_{\sigma \sigma
^{\prime }}(\tau -\tau ^{\prime })$ is depicted on the $Fig.6$

The final equations for renormalized functions it is more
convenient to write down in Fourier representation
%
%%%%%%%%%%%%%%%%
\begin{widetext}
%%%%%%%%%%%%%%%%
%
%%%%%%%%%%%%%%%%
\begin{eqnarray}
%%%%%%%%%%%%%%%%
%
G(\mathbf{k,}\sigma ,\tau |\mathbf{k}^{\prime },\sigma ^{\prime
},\tau ^{\prime }) &=&\frac{1}{\beta }\sum\limits_{\omega
}G_{\sigma \sigma ^{\prime }}(\mathbf{k},\mathbf{k}^{\prime
}|\,i\omega )\,\exp \left[
-i\omega (\tau -\tau ^{^{\prime }})\right] ,  \nonumber \\
F(\mathbf{k,}\sigma ,\tau |-\mathbf{k}^{\prime },-\sigma ^{\prime
},\tau ^{\prime }) &=&\frac{1}{\beta }\sum\limits_{\omega
}F_{\sigma \overline{ \sigma }^{\prime
}}(\mathbf{k},-\mathbf{k}^{\prime }|\,i\omega )\exp \left[
-i\omega (\tau -\tau ^{^{\prime }})\right] . \nonumber
%
%%%%%%%%%%%%%%%
\end{eqnarray}
%%%%%%%%%%%%%%
%
The complete equations for the conduction electrons propagators
have the form:
%
%%%%%%%%%%%%%%%%
\begin{eqnarray}
%%%%%%%%%%%%%%%%
%
G_{\sigma \sigma ^{\prime }}(\mathbf{k},\mathbf{k}^{\prime
}|\,i\omega ) &=&\delta _{\mathbf{kk}^{\prime }}\delta _{\sigma
\sigma ^{\prime }}G_{\sigma }^{0}(\mathbf{k}|\,i\omega
)+\frac{V_{\mathbf{k}}^{\ast }V_{ \mathbf{k}^{\prime
}}}{N}(G_{\sigma }^{0}(\mathbf{k}|i\omega )g_{\sigma \sigma
^{\prime }}(i\omega )G_{\sigma ^{\prime }}^{0}(\mathbf{k}^{\prime
}|i\omega )-  \nonumber \\
&-&G_{\sigma }^{0}(\mathbf{k}|i\omega )f_{\sigma \overline{\sigma
}^{\prime }}(i\omega )\overline{F}_{\overline{\sigma }^{\prime
}\sigma ^{\prime }}^{0}(-\mathbf{k}^{\prime }|i\omega ))-F_{\sigma
\overline{\sigma }}^{0}( \mathbf{k}|i\omega )g_{\overline{\sigma
}^{\prime } \overline{\sigma }}(-i\omega
)\overline{F}_{\overline{\sigma }^{\prime }\sigma ^{\prime
}}^{0}(-\mathbf{k}|i\omega )-  \nonumber \\
&-&F_{\sigma \overline{\sigma }}^{0}(\mathbf{k}|i\omega
)\overline{f}_{ \overline{\sigma }\sigma ^{\prime }}(i\omega
)G_{\sigma ^{\prime }}^{0}(
\mathbf{k}^{\prime }|i\omega )),\label{16} \\
F_{\sigma \overline{\sigma }^{\prime
}}(\mathbf{k},-\mathbf{k}^{\prime }|\,i\omega ) &=&F_{\sigma
\overline{\sigma }}^{0}(\mathbf{k}|\,i\omega )\delta
_{\mathbf{kk}^{\prime }}\delta _{\sigma \sigma ^{\prime
}}+\frac{V_{ \mathbf{k}}^{\ast }V_{\mathbf{k}^{\prime
}}}{N}(G_{\sigma }^{0}(\mathbf{k} |i\omega )g_{\sigma \sigma
^{\prime }}(i\omega )F_{\sigma ^{\prime } \overline{\sigma
}^{\prime }}^{0}(\mathbf{k}^{\prime }|i\omega )+ \nonumber
\\
&+&G_{\sigma }^{0}(\mathbf{k}|i\omega )f_{\sigma \overline{\sigma
}^{\prime }}(i\omega )G_{\overline{\sigma }^{\prime
}}^{0}(-\mathbf{k}^{\prime }|-i\omega )+F_{\sigma \overline{\sigma
}}^{0}(\mathbf{k}|i\omega )g_{ \overline{\sigma }^{\prime
}\overline{\sigma }}(-i\omega )G_{\overline{
\sigma }^{\prime }}^{0}(-\mathbf{k}^{\prime }|-i\omega )-  \nonumber \\
&-&F_{\sigma \overline{\sigma }}^{0}(\mathbf{k}|i\omega
)\overline{f}_{ \overline{\sigma }\sigma ^{\prime }}(i\omega
)F_{\sigma ^{\prime }\overline{ \sigma }^{\prime
}}^{0}(\mathbf{k}^{\prime }|i\omega )).\label{17}
%
%%%%%%%%%%%%%%
\end{eqnarray}
%%%%%%%%%%%%%%
%
%%%%%%%%%%%%%%
\end{widetext}
%%%%%%%%%%%%%%
%
\begin{figure*}[t]
%%%%%%%%%%%%%%%%%%
%
\centering
\includegraphics[width=0.95\textwidth,clip]{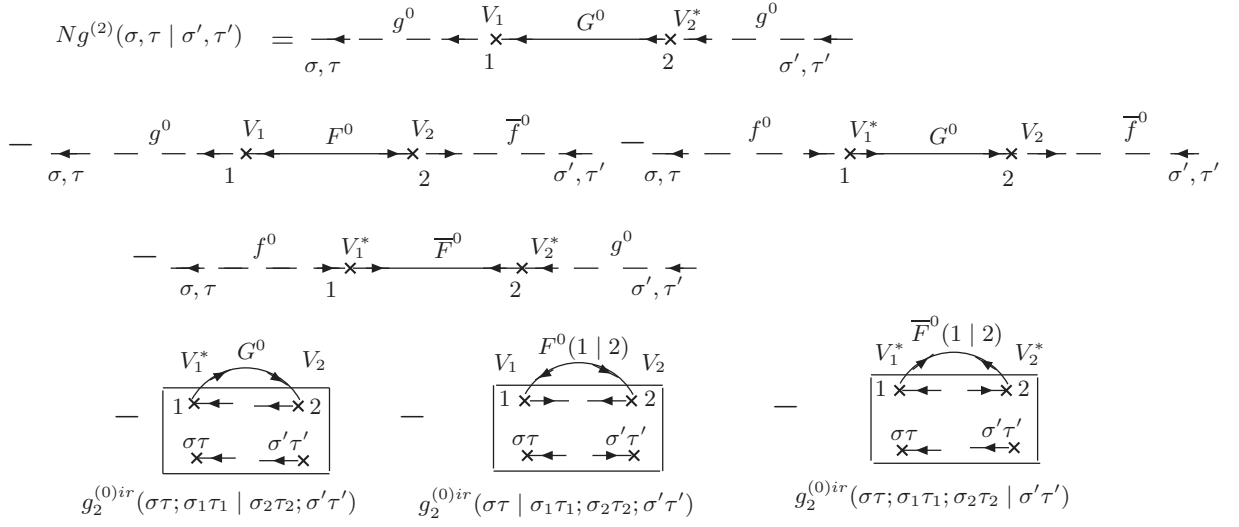}
\caption{ The second order perturbation contribution for the
impurity electron normal propagator.}\label{fig-5}
%
%%%%%%%%%%%%%
\end{figure*}
%%%%%%%%%%%%%
%
\begin{figure*}[t]
%%%%%%%%%%%%%%%%%%
%
\centering
\includegraphics[width=0.95\textwidth,clip]{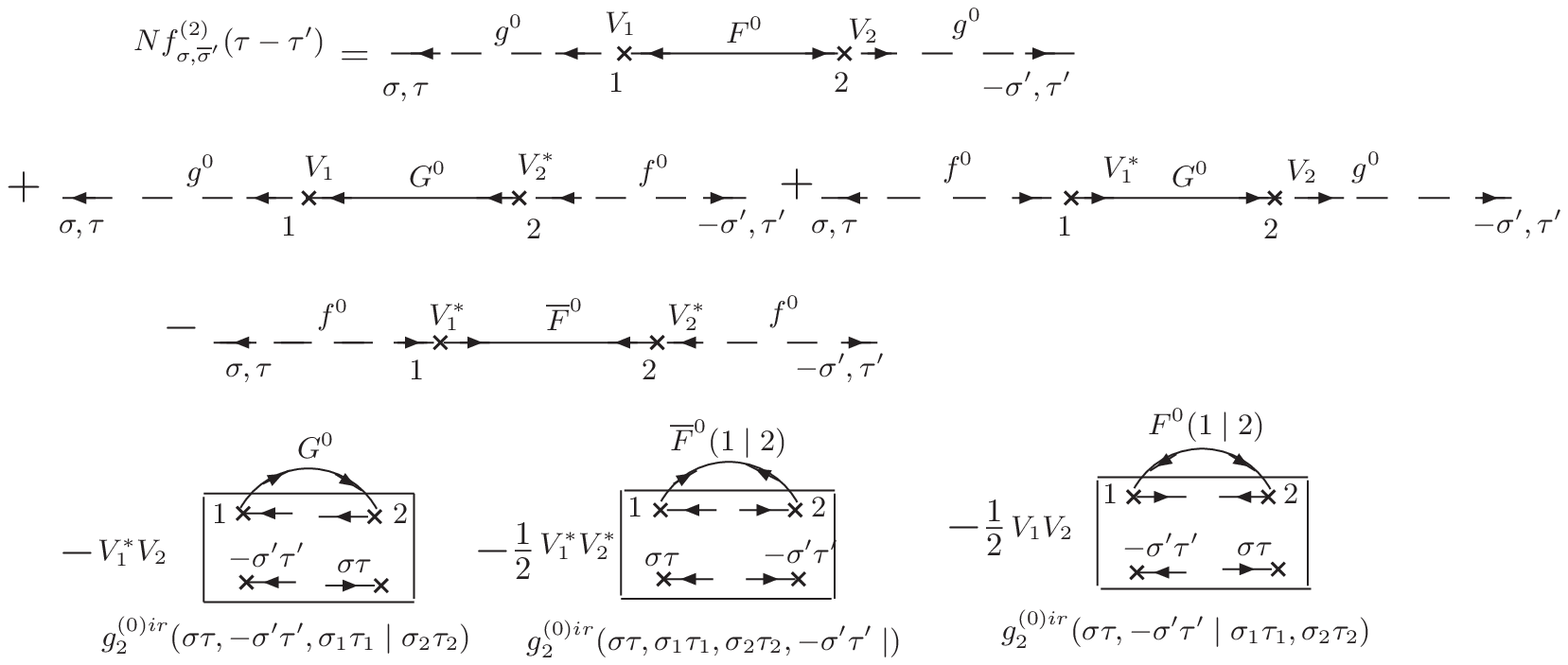}
\caption{ Anomalous impurity electron Green's function in the
second order perturbation theory. }\label{fig-6}
%
%%%%%%%%%%%%%
\end{figure*}
%%%%%%%%%%%%%
%
These renormalized propagators are expressed through the full
propagators $ g,f$ and $\overline{f}$\ of impurity electrons. Now
it is necessary to obtain the corresponding equations for the full
impurity electron propagators. Because the subsystem of
$f$-electrons is strongly correlated we have to introduce the
correlation functions $Z_{\sigma \sigma ^{\prime }},Y_{\sigma
\overline{\sigma }^{\prime }}$ and $\overline{Y}_{\overline{
\sigma }\sigma ^{\prime }}$ which are represented by strong
connected diagrams with irreducible Green's functions$^{[31-35]}$.
The process of renormalization of $f$-electron propagators is
shown on the $Figures$ $7$ and $8$, where the double dashed lines
depict the full $f$-electron functions and the rectangles
represent the correlation functions $\Lambda _{\sigma \sigma
^{\prime }}=g_{\sigma \sigma ^{\prime }}^{0}+Z_{\sigma \sigma
^{\prime }}$, $Y_{\sigma \sigma ^{\prime }}$ and $\overline{Y}_{
\overline{\sigma }\sigma }$:
\begin{figure*}[t]
%%%%%%%%%%%%%%%%%%
%
\centering
\includegraphics[width=0.95\textwidth,clip]{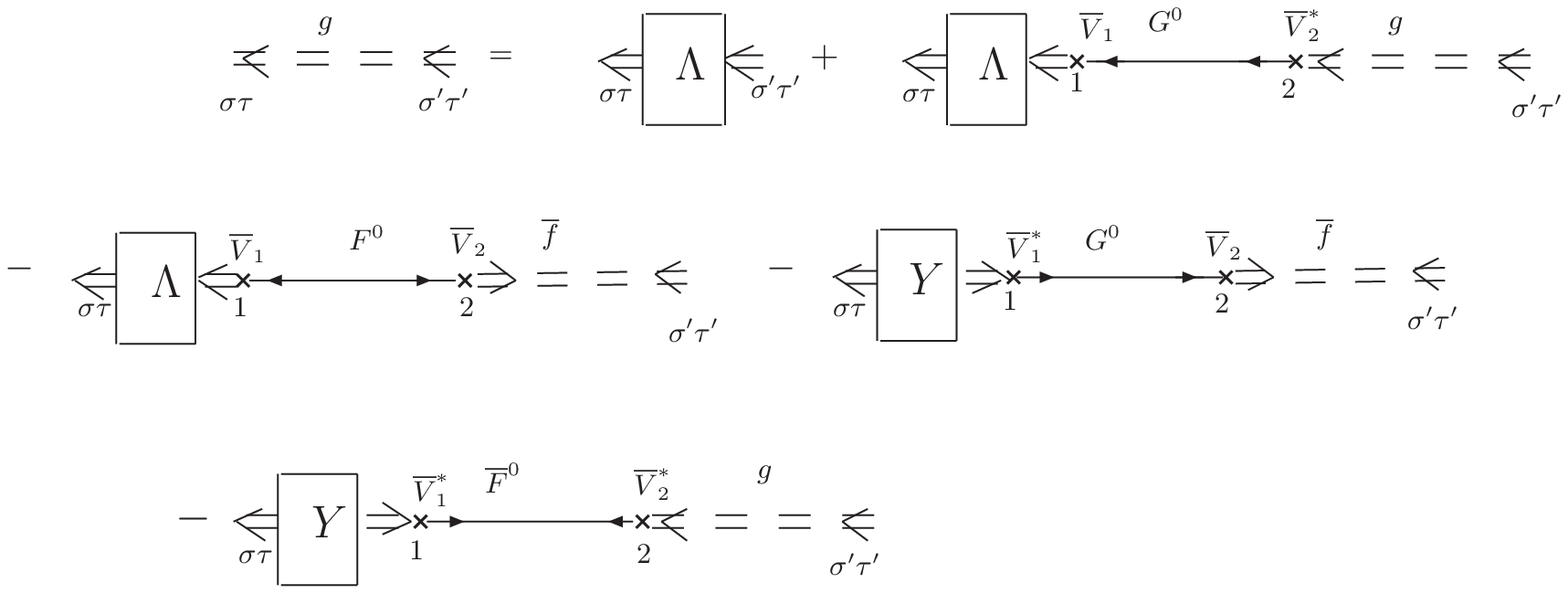}
\caption{ Dyson type equation for the normal propagator of
impurity electrons. Double dashed lines depict full electron
propagators. The arrows on them distinguish the normal and
anomalous ones. The squares with attached arrows depict the
correlated functions. On double repeated indices 1 and 2 is
supposed summation by $\sigma _{n}$ and $\mathbf{k}_{n}$ and
integration by \ $\tau _{n}$\ .}\label{fig-7}
%
%%%%%%%%%%%%%
\end{figure*}
%%%%%%%%%%%%%
%
The second equation we shall depict for anomalous propagator
$\overline{f}$ of the impurity electrons (see $Fig.8$).
\begin{figure*}[t]
%%%%%%%%%%%%%%%%%%
%
\centering
\includegraphics[width=0.95\textwidth,clip]{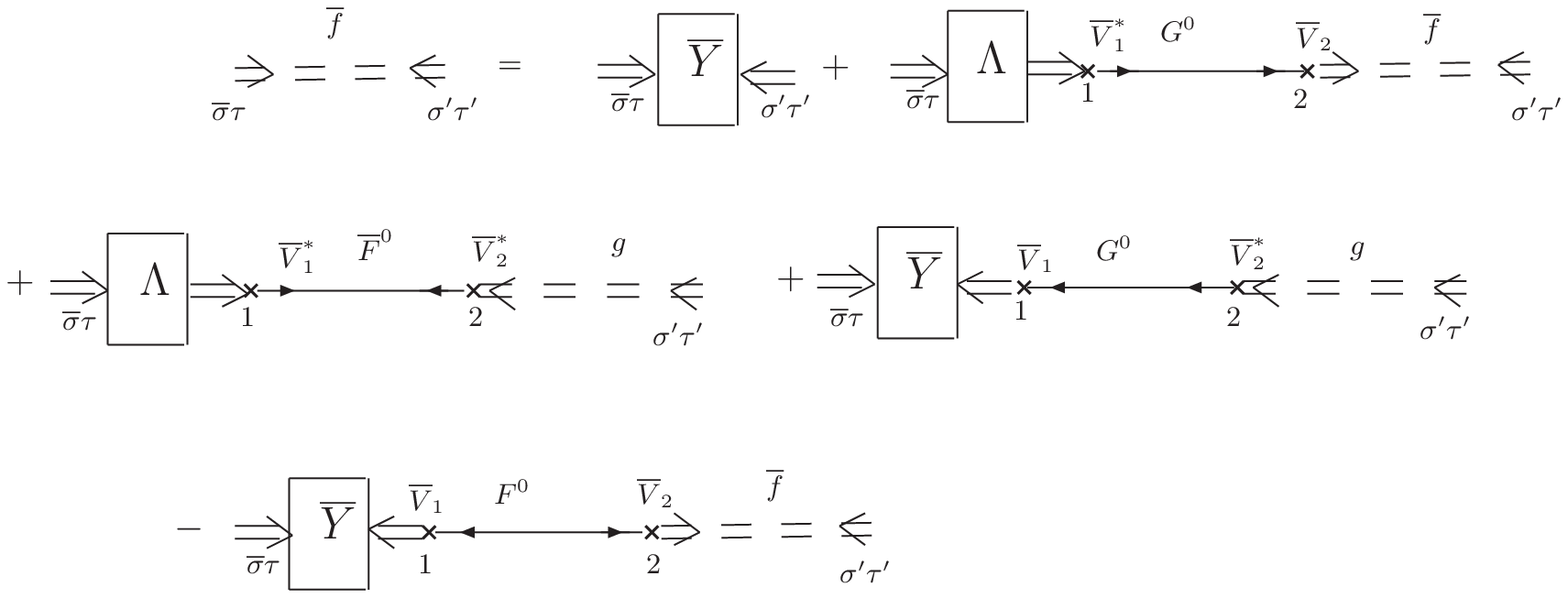}
\caption{ Dyson type equation for one of anomalous Green's
functions of $f$ -electrons. }\label{fig-8} \vspace{-5mm}
%
%%%%%%%%%%%%%
\end{figure*}
%%%%%%%%%%%%%
%
In both these equations the bare conduction electron propagators
$G_{\sigma }^{0}(\mathbf{k}|i\omega )$, $F_{\sigma
\overline{\sigma }}^{0}(\mathbf{k} |i\omega )$ and \newline
$\overline{F}_{\overline{\sigma }\sigma }^{0}(-\mathbf{k}|i\omega
)$ play the role of mass operators for the $f$-electron
propagators. It is easy to see that these functions participate in
above equations being averaged on the Brillouin cell with matrix
elements of hybridization. Therefore we define the new quantities
\begin{figure*}[t]
%%%%%%%%%%%%%%%%%%
%
\centering
\includegraphics[width=0.65\textwidth,clip]{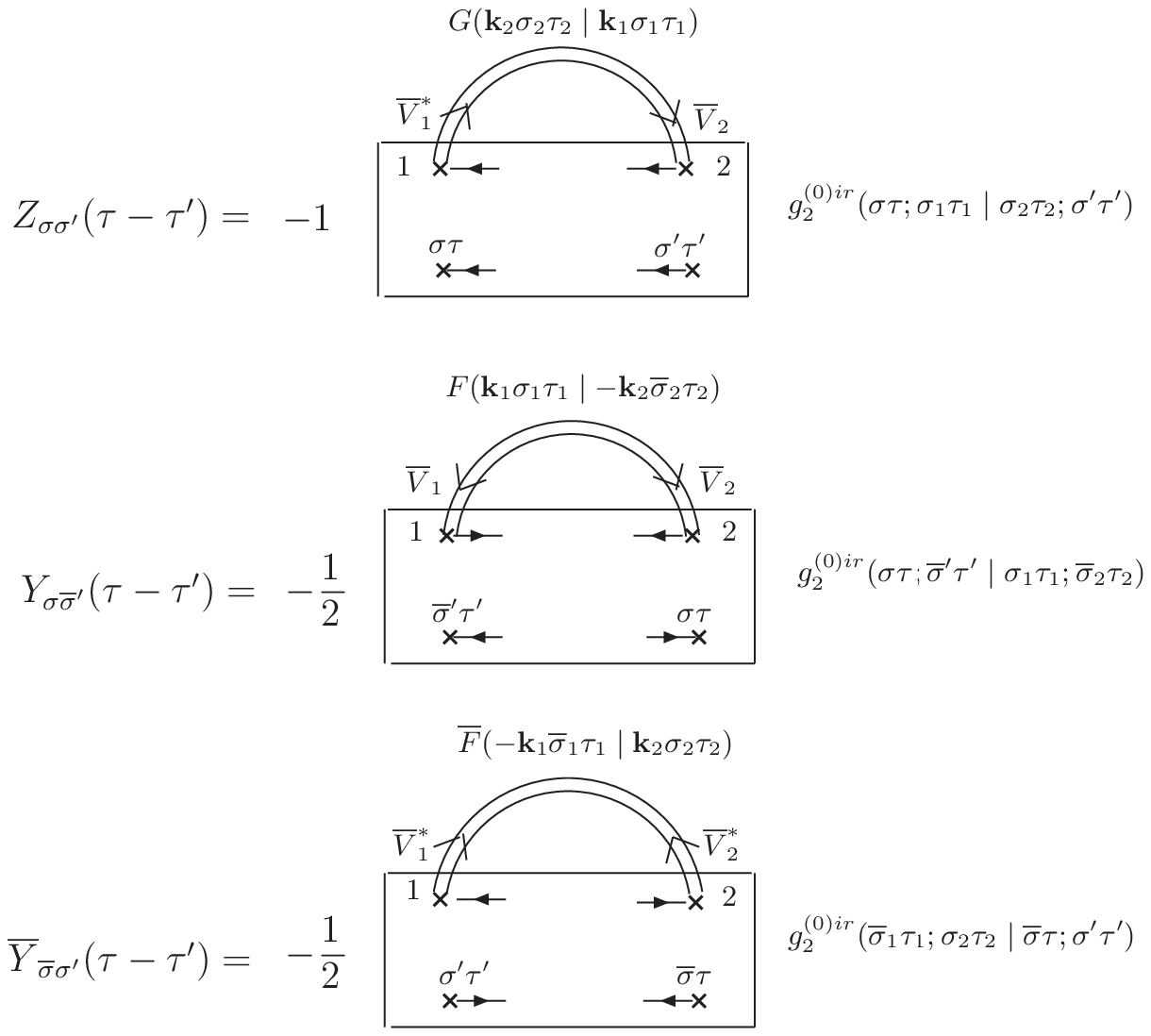}
\caption{ Schematic representation of the main approximations for
the correlated functions The solid double lines with arrows depict
the renormalized one-particle Green's functions of conduction
electrons. The rectangles depict the irreducible Green's functions
of impurity electrons.}\label{fig-9}
\vspace{-5mm}
%
%%%%%%%%%%%%%
\end{figure*}
%%%%%%%%%%%%%
%
\begin{figure*}[t]
%%%%%%%%%%%%%%%%%%
%
\centering
\includegraphics[width=0.85\textwidth,clip]{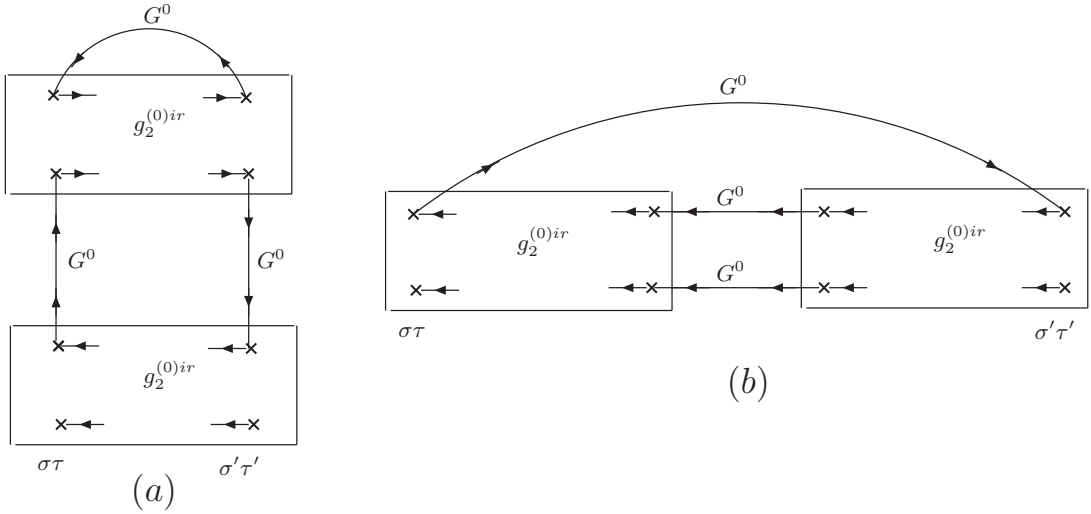}
\caption{ Examples of the simplest ladder diagram for
$g$-function.}\label{fig-10}
\vspace{-5mm}
%
%%%%%%%%%%%%%
\end{figure*}
%%%%%%%%%%%%%
%
%%%%%%%%%%%%%%%%
\begin{widetext}
%%%%%%%%%%%%%%%%
%
%%%%%%%%%%%%%%%%
\begin{eqnarray}
%%%%%%%%%%%%%%%%
%
\hspace{-0.7cm}\frac{1}{N}\sum\limits_{\mathbf{k}_{1}\mathbf{k}_{2}}V_{
\mathbf{k}_{2}}^{\ast
}V_{\mathbf{k}_{1}}G^{0}(\mathbf{k}_{1},\sigma _{1},\tau
_{1}|\mathbf{k}_{2},\sigma _{2},\tau _{2}) &=&\frac{1}{N}
\sum\limits_{\mathbf{k}_{1}}\,|V_{\mathbf{k}_{1}}|^{2}G_{\sigma
_{1}\sigma _{2}}^{0}(\mathbf{k}|\tau _{1}-\tau _{2})\equiv \delta
_{\sigma _{1}\sigma
_{2}}G_{\sigma _{1}}^{0}(\tau _{1}-\tau _{2}),  \nonumber \\
\hspace{-0.7cm}\frac{1}{N}\sum\limits_{\mathbf{k}_{1}\mathbf{k}_{2}}V_{
\mathbf{k}_{1}}^{\ast }V_{\mathbf{k}_{2}}^{\ast
}\overline{F}^{0}(\mathbf{k} _{1},\overline{\sigma }_{1},\tau
_{1}|\mathbf{k}_{2},\sigma _{2},\tau _{2})
&=&\frac{1}{N}\sum\limits_{\mathbf{k}_{1}}\,|V_{\mathbf{k}_{1}}|^{2}\overline{F}_{
\overline{\sigma }_{1}\sigma _{2}}^{0}(-\mathbf{k}_{1}|\tau
_{1}-\tau _{2})\equiv \delta _{\sigma _{1}\sigma
_{2}}\overline{F}_{\overline{\sigma }
_{1}\sigma _{1}}^{0}(\tau _{1}-\tau _{2}), \\
\hspace{-0.7cm}\frac{1}{N}\sum\limits_{\mathbf{k}_{1}\mathbf{k}_{2}}V_{
\mathbf{k}_{1}}V_{\mathbf{k}_{2}}F^{0}(\mathbf{k}_{1},\sigma
_{1},\tau _{1}| \mathbf{k}_{2},\overline{\sigma }_{2},\tau _{2})
&=&\frac{1}{N}\sum\limits_{
\mathbf{k}_{1}}\,|V_{\mathbf{k}_{1}}|^{2}F_{\sigma
_{1}\overline{\sigma } _{2}}^{0}(\mathbf{k}_{1}|\tau _{1}-\tau
_{2})\equiv \delta _{\sigma _{1}\sigma _{2}}F_{\sigma
_{1}\overline{\sigma }_{1}}^{0}(\tau _{1}-\tau _{2}).  \nonumber
\label{18}
%
%%%%%%%%%%%%%%
\end{eqnarray}
%%%%%%%%%%%%%%
%
These definitions gives us the possibility to simplify the
structure of equations for the $f$-electron propagators. By using
these average bare propagators $G_{\sigma }^{0}$, $F_{\sigma
\overline{\sigma }}^{0}$ and $ \overline{F}_{\overline{\sigma
}\sigma }^{0}$ and Fourier representation for $\tau $-variables we
obtain
%
%%%%%%%%%%%%%%%%
\begin{eqnarray}
%%%%%%%%%%%%%%%%
%
g_{\sigma }(i\omega ) & = & \frac{\Lambda _{\sigma }(i\omega
)\,\,-\,\,G_{ \overline{\sigma }}^{0}(-i\omega )[\Lambda _{\sigma
}(i\omega )\Lambda _{ \overline{\sigma }}(-i\omega
)\,\,+\,\,Y_{\sigma \overline{\sigma }}(i\omega
)\overline{Y}_{\overline{\sigma }\sigma }(i\omega )]}{d_{\sigma
}(i\omega )},\label{19}
\\
\overline{f}_{\overline{\sigma }\sigma }(i\omega ) & = &
\frac{\overline{Y}_{ \overline{\sigma }\sigma }(i\omega
)\,\,+\,\,\overline{F}_{\overline{\sigma } \sigma }^{0}(i\omega
)(\Lambda _{\sigma }(i\omega )\Lambda _{\overline{ \sigma
}}(-i\omega )\,\,+\,\,Y_{\sigma \overline{\sigma }}(i\omega )
\overline{Y}_{\overline{\sigma }\sigma }(i\omega ))}{d_{\sigma
}(i\omega )},\label{20}
\\
f_{\sigma \overline{\sigma }}(i\omega ) & = & \frac{\{Y_{\sigma
\overline{ \sigma }}(i\omega )\,\,+\,F_{\sigma \overline{\sigma
}}^{0}(i\omega )[(\Lambda _{\sigma }(i\omega )\Lambda
_{\overline{\sigma }}(-i\omega )\,\,+\,\,Y_{\sigma
\overline{\sigma }}(i\omega )\overline{Y}_{\overline{
\sigma }\sigma }(i\omega )]\}}{d_{\sigma }(i\omega )},\label{21} \\
d_{\sigma }(i\omega ) &=&(1-\Lambda _{\sigma }(i\omega )G_{\sigma
}^{0}(i\omega ))(1-\Lambda _{\overline{\sigma }}(-i\omega
)G_{\overline{ \sigma }}^{0}(-i\omega
))+\overline{Y}_{\overline{\sigma }\sigma }(i\omega
)F_{\sigma \overline{\sigma }}^{0}(i\omega )+  \nonumber \\
&+&Y_{\sigma \overline{\sigma }}(i\omega
)\overline{F}_{\overline{\sigma } \sigma }^{0}(i\omega
)+\overline{F}_{\overline{\sigma }\sigma }^{0}(i\omega )F_{\sigma
\overline{\sigma }}^{0}(i\omega )[Y_{\sigma \overline{\sigma }
}(i\omega )\overline{Y}_{\overline{\sigma }\sigma }(i\omega
)+\Lambda
_{\sigma }(i\omega )\Lambda _{\overline{\sigma }}(-i\omega )]+  \nonumber \\
&+&G_{\overline{\sigma }}^{0}(-i\omega )G_{\sigma }^{0}(i\omega
)Y_{\sigma \overline{\sigma }}(i\omega
)\overline{Y}_{\overline{\sigma }\sigma }(i\omega ).\label{22}
%
%%%%%%%%%%%%%%
\end{eqnarray}
%%%%%%%%%%%%%%
%

In the previous part of the paper we supposed the existence of
pairing potential of conduction electrons with order parameter and
with the bare propagators:
%
%%%%%%%%%%%%%%%%
\begin{eqnarray}
%%%%%%%%%%%%%%%%
%
G_{\sigma }^{0}(\mathbf{k}|i\omega )& = & \frac{i\omega
\,+\,\epsilon (\mathbf{k} ) }{(i\omega )^{2}-E^{2}(\mathbf{k})};
\,\, F_{\sigma \overline{\sigma } }^{0}( \mathbf{k}|i\omega
)=\overline{F}_{\overline{\sigma }\sigma }^{0}(- \mathbf{k
}|i\omega )=\frac{\Delta }{(i\omega )^{2}-\,E^{2}(\mathbf{k})};
\,\, E( \mathbf{k})=\sqrt{\epsilon ^{2}(\mathbf{k})+\Delta
^{2}}.\label{23}
%
%%%%%%%%%%%%%%
\end{eqnarray}
%%%%%%%%%%%%%%
%
Now we shall discuss the case when the pairing potential is absent
and the superconducting state appears simultaneously \ with both
subsystems as a consequence of the broken symmetry and phase
transition. In this more simple case the renormalized conduction
electron propagators have the form
%
%%%%%%%%%%%%%%%%
\begin{eqnarray}
%%%%%%%%%%%%%%%%
%
G_{\sigma \sigma ^{\prime }}(\mathbf{k},\mathbf{k}|i\omega )& = &
\delta _{ \mathbf{kk}}\delta _{\sigma \sigma ^{\prime }}G_{\sigma
}^{0}(\mathbf{k} |i\omega )+\frac{V_{\mathbf{k}}^{\ast
}V_{\mathbf{k}}}{N}G_{\sigma }^{0}( \mathbf{k}|i\omega )g_{\sigma
\sigma ^{\prime }}(i\omega )G_{\sigma ^{\prime
}}^{0}(\mathbf{k}|i\omega ), \label{24}\\
F_{\sigma \overline{\sigma }^{\prime
}}(\mathbf{k},-\mathbf{k}|i\omega ) & =&
\frac{V_{\mathbf{k}}^{\ast }V_{\mathbf{k}}}{N}G_{\sigma
}^{0}(\mathbf{k} |i\omega )f_{\sigma \overline{\sigma }^{\prime
}}(i\omega )G_{\overline{
\sigma }^{\prime }}^{0}(-\mathbf{k}|-i\omega ), \label{25}\\
G_{\sigma }^{0}(\mathbf{k}|i\omega ) & = & (i\omega -\epsilon
(\mathbf{k} ))^{-1}.\label{26}
%
%%%%%%%%%%%%%%
\end{eqnarray}
%%%%%%%%%%%%%%
%

The renormalized propagators of impurity electron in this case
are:
%
%%%%%%%%%%%%%%%%
\begin{eqnarray}
%%%%%%%%%%%%%%%%
%
g_{\sigma }(i\omega )& = & \frac{\Lambda _{\sigma }(i\omega
)\,\,-\,\,G_{ \overline{\sigma }}^{0}(-i\omega )[\Lambda _{\sigma
}(i\omega )\Lambda _{ \overline{\sigma }}(-i\omega
)\,\,+\,\,Y_{\sigma \overline{\sigma }}(i\omega
)\overline{Y}_{\overline{\sigma }\sigma }(i\omega )]}{d_{\sigma
}(i\omega )},\label{27}
\\
\overline{f}_{\overline{\sigma }\sigma }(i\omega ) &
=&\frac{\overline{Y}_{ \overline{\sigma }\sigma }(i\omega
)\,\,}{d_{\sigma }(i\omega )};\,\, f_{\sigma \overline{\sigma
}}(i\omega ) = \frac{Y_{\sigma \overline{\sigma }
}(i\omega )\,\,}{d_{\sigma }(i\omega )},\label{28} \\
d_{\sigma }(i\omega ) & = & (1-\Lambda _{\sigma }(i\omega
)G_{\sigma }^{0}(i\omega ))(1-\Lambda _{\overline{\sigma
}}(-i\omega)G_{\overline{
\sigma }}^{0}(-i\omega ))+  \nonumber \\
& + & G_{\overline{\sigma }}(-i\omega )G_{\sigma
}^{0}(i\omega)Y_{\sigma \overline{\sigma }}(i\omega
)\overline{Y}_{\overline{\sigma }\sigma }(i\omega ).\label{29}
%
%%%%%%%%%%%%%%
\end{eqnarray}
%%%%%%%%%%%%%%
%
%%%%%%%%%%%%%%
\end{widetext}
%%%%%%%%%%%%%%
%
The equation (24) has been established many years ago in the paper
of Anderson$^{[1]}$ by using the equation of motion of conduction
electron operators. In this equation the propagator $g_{\sigma
}(i\omega )$ has the role of $t$-matrix for non-spin-flip
scattering. By setting $\mathbf{k}= \mathbf{k}^{\prime }$ in
$G_{\sigma }(\mathbf{k},\mathbf{k}^{\prime }|i\omega )$ \newline
%
%%%%%%%%%%%%%%%%
\begin{equation}
%%%%%%%%%%%%%%%%
%
G_{\sigma }(\mathbf{k},\mathbf{k}^{\prime }|i\omega
)=\frac{1}{i\omega
\,-\,\epsilon (\mathbf{k})}+\frac{|V_{\mathbf{k}}|^{2}g_{\sigma }(i\omega )}{%
N(i\omega \,-\,\epsilon (\mathbf{k}))^{2}}\label{30}
%
%%%%%%%%%%%%%%
\end{equation}
%%%%%%%%%%%%%%
%
and considering the Lehmann spectral representation it is possible
to conclude that the discontinuity of $g_{\sigma }(E)$ across the
real axis is pure imaginary$^{[8]}$
%
%%%%%%%%%%%%%%%%
\begin{equation}
%%%%%%%%%%%%%%%%
%
g_{\sigma }(E+i\delta )=[g_{\sigma }(E-i\delta )]^{\ast
}.\label{31}
%
%%%%%%%%%%%%%%
\end{equation}
%%%%%%%%%%%%%%
%

The Green's function $g_{\sigma }(i\omega )$ has been known till
now in approximate form as a result of special decoupling
mechanism used for equation of motion of quantum Green's
functions. As is known in such decoupling approximation some
combinations of operators is taken off the average value of
product of operators and are replaced by their average values.
After that truncation the Green's functions of low order remain.
This approximation has been proposed by Bogoliubov, Tiablikov,
Zubarev and Tserkovnikov$^{[21-24]}$ and used by other
authors$^{[2-14,18]}$. The hybridization of conduction and
impurity electrons causes the appearance of mixed Green's
functions:
%
%%%%%%%%%%%%%%%%
\begin{eqnarray}
%%%%%%%%%%%%%%%%
%
G_{m}(\mathbf{k},\sigma ,\tau |\sigma ^{\prime },\tau ^{\prime })
&=&-\left\langle TC_{\mathbf{k}\sigma }(\tau )\overline{f}_{\sigma
^{\prime
}}(\tau ^{\prime })U(\beta )\right\rangle _{0}^{c},  \nonumber \\
F_{m}(\mathbf{k},\sigma ,\tau |\overline{\sigma }^{\prime },\tau ^{\prime })
&=&-\left\langle TC_{\mathbf{k}\sigma }(\tau )f_{\overline{\sigma }^{\prime
}}(\tau ^{\prime })U(\beta )\right\rangle _{0}^{c},\qquad \label{32}\\
\overline{F}_{m}(-\mathbf{k},\overline{\sigma },\tau |\sigma
^{\prime },\tau ^{\prime }) &=&-\left\langle
T\overline{C}_{-\mathbf{k}\overline{\sigma } }(\tau
)\overline{f}_{\sigma ^{\prime }}(\tau ^{\prime })U(\beta
)\right\rangle _{0}^{c},  \nonumber
%
%%%%%%%%%%%%%%
\end{eqnarray}
%%%%%%%%%%%%%%
%
and also
%
%%%%%%%%%%%%%%%%
\begin{eqnarray}
%%%%%%%%%%%%%%%%
%
G^{m}(\sigma ,\tau |\mathbf{k},\sigma ^{\prime },\tau ^{\prime })
&=&-\left\langle Tf_{\sigma }(\tau )\overline{C}_{\mathbf{k}\sigma
^{\prime
}}(\tau ^{\prime })U(\beta )\right\rangle _{0}^{c},  \nonumber \\
F^{m}(\sigma ,\tau |-\mathbf{k},\overline{\sigma }^{\prime },\tau
^{\prime }) &=&-\left\langle Tf_{\sigma }(\tau
)C_{-\mathbf{k}\overline{\sigma }
^{\prime }}(\tau ^{\prime })U(\beta )\right\rangle _{0}^{c},\label{33} \\
\overline{F}^{m}(\overline{\sigma },\tau |\mathbf{k},\sigma
^{\prime },\tau ^{\prime }) &=&-\left\langle
T\overline{f}_{\overline{\sigma }}(\tau )
\overline{C}_{\mathbf{k}\sigma ^{\prime }}(\tau ^{\prime })U(\beta
)\right\rangle _{0}^{c}.  \nonumber
%
%%%%%%%%%%%%%%
\end{eqnarray}
%%%%%%%%%%%%%%
%
Let $G_{m\sigma \sigma ^{\prime }}(\mathbf{k}|i\omega )$,
$F_{m\sigma \overline{\sigma }^{\prime }}(\mathbf{k}|i\omega )$
and $F_{m\overline{ \sigma }\sigma ^{\prime }}(\mathbf{k}|i\omega
)$ be the Fourier representation of the first group of Green's
functions and $G_{\sigma \sigma ^{\prime }}^{m}(\mathbf{k}|i\omega
)$, $F_{\sigma \overline{\sigma }^{\prime
}}^{m}(-\mathbf{k}|i\omega )$ and $\overline{F}_{\overline{\sigma
}\sigma ^{\prime }}^{m}(\mathbf{k}|i\omega )$ of the second
group.\newline In the presence of superconducting pairing of
conduction electrons we obtain the following results:
%
%%%%%%%%%%%%%%%%
\begin{widetext}
%%%%%%%%%%%%%%%%
%
%%%%%%%%%%%%%%%%
\begin{eqnarray}
%%%%%%%%%%%%%%%%
%
G_{m\sigma \sigma ^{\prime }}(\mathbf{k}|i\omega )
&=&\frac{V_{\mathbf{k} }^{\ast }}{\sqrt{N}}\left[ G_{\sigma
}^{0}(\mathbf{k}|i\omega )g_{\sigma \sigma ^{\prime }}(i\omega
)-F_{\sigma \overline{\sigma }}^{0}(\mathbf{k} |i\omega
)\overline{f}_{\overline{\sigma }\sigma ^{\prime }}(i\omega
)\right]
,  \nonumber \\
F_{m\sigma \overline{\sigma }^{\prime }}(\mathbf{k}|i\omega )
&=&\frac{V_{ \mathbf{k}}^{\ast }}{\sqrt{N}}\left[ G_{\sigma
}^{0}(\mathbf{k}|i\omega )f_{\sigma \sigma ^{\prime }}(i\omega
)+F_{\sigma \overline{\sigma }^{\prime }}^{0}(\mathbf{k}|i\omega
)g_{\overline{\sigma }^{\prime }\overline{\sigma }
}(-i\omega )\right] ,\label{34} \\
\overline{F}_{m\overline{\sigma }\sigma ^{\prime
}}(-\mathbf{k}|i\omega ) &=& \frac{V_{\mathbf{k}}^{\ast
}}{\sqrt{N}}\left[ G_{\overline{\sigma }}^{0}(-
\mathbf{k}|-i\omega )\overline{f}_{\overline{\sigma }\sigma
^{\prime }}(i\omega )+\overline{F}_{\overline{\sigma }\sigma
}^{0}(-\mathbf{k} |i\omega )g_{\sigma \sigma ^{\prime }}(i\omega
)\right] . \nonumber
%
%%%%%%%%%%%%%%
\end{eqnarray}
%%%%%%%%%%%%%%
%
For the second group of mixed propagators we have:
%
%%%%%%%%%%%%%%%%
\begin{eqnarray}
%%%%%%%%%%%%%%%%
%
G_{\sigma \sigma ^{\prime }}^{m}(\mathbf{k}|i\omega )
&=&\frac{V_{\mathbf{k}} }{\sqrt{N}}\left[ g_{\sigma \sigma
^{\prime }}(i\omega )G_{\sigma ^{\prime }}^{0}(\mathbf{k}|i\omega
)-f_{\sigma \overline{\sigma }^{\prime }}(i\omega )
\overline{F}_{\overline{\sigma }^{\prime }\sigma ^{\prime
}}^{0}(-\mathbf{k}
|i\omega )\right] ,  \nonumber \\
F_{\sigma \overline{\sigma }^{\prime }}^{m}(-\mathbf{k}|i\omega )
&=&\frac{ V_{\mathbf{k}}}{\sqrt{N}}\left[ g_{\sigma \sigma
^{\prime }}(i\omega )F_{\sigma ^{\prime }\overline{\sigma
}^{\prime }}^{0}(\mathbf{k}|i\omega )+f_{\sigma \overline{\sigma
}^{\prime }}(i\omega )G_{\overline{\sigma }
^{\prime }}^{0}(-\mathbf{k}|i\omega )\right] ,\label{35} \\
\overline{F}_{\overline{\sigma }\sigma ^{\prime
}}^{m}(\mathbf{k}|i\omega )
&=&\frac{V_{\mathbf{k}}}{\sqrt{N}}\left[
\overline{f}_{\overline{\sigma } \sigma ^{\prime }}(i\omega
)G_{\sigma ^{\prime }}^{0}(\mathbf{k}|i\omega
)+g_{\overline{\sigma }\sigma ^{\prime }}(-i\omega
)\overline{F}_{\overline{ \sigma }^{\prime }\sigma ^{\prime
}}^{0}(-\mathbf{k}|i\omega )\right] . \nonumber
%
%%%%%%%%%%%%%%
\end{eqnarray}
%%%%%%%%%%%%%%
%

Now we multiply the system of operators (33) by
$V_{\mathbf{k}}^{\ast }/ \sqrt{N}$ and sum after $\mathbf{k}$, use
the definitions (18) and suppose the paramagnetic phase of the
system. Then we obtain:
%
%%%%%%%%%%%%%%%%
\begin{eqnarray}
%%%%%%%%%%%%%%%%
%
G_{\sigma }^{m}(i\omega )
&=&\frac{1}{\sqrt{N}}\sum\limits_{\mathbf{k}}V_{ \mathbf{k}}^{\ast
}G_{\sigma }^{m}(\mathbf{k}|i\omega )=g_{\sigma }(i\omega
)G_{\sigma }^{0}(i\omega )-f_{\sigma \overline{\sigma }^{\prime
}}(i\omega )
\overline{F}_{\overline{\sigma }\sigma }^{0}(i\omega ),  \nonumber \\
F_{\sigma \overline{\sigma }}^{m}(i\omega ) &=&\frac{1}{\sqrt{N}}
\sum\limits_{\mathbf{k}}V_{\mathbf{k}}^{\ast }F_{\sigma
\overline{\sigma } }^{m}(-\mathbf{k}|i\omega )=g_{\sigma }(i\omega
)F_{\sigma \overline{\sigma } }^{0}(i\omega )+f_{\sigma
\overline{\sigma }}(i\omega )G_{\overline{\sigma }
}^{0}(-i\omega ),\label{36} \\
\overline{F}_{\overline{\sigma }\sigma }^{m}(i\omega )
&=&\frac{1}{\sqrt{N}} \sum\limits_{\mathbf{k}}V_{\mathbf{k}}^{\ast
}\overline{F}_{\overline{\sigma }\sigma }^{m}(\mathbf{k}|i\omega
)=g_{\overline{\sigma }}(-i\omega ) \overline{F}_{\overline{\sigma
}\sigma }^{0}(i\omega )+\overline{f}_{ \overline{\sigma }\sigma
}(i\omega )G_{\sigma }^{0}(i\omega ). \nonumber
%
%%%%%%%%%%%%%%
\end{eqnarray}
%%%%%%%%%%%%%%
%
%%%%%%%%%%%%%%
\end{widetext}
%%%%%%%%%%%%%%
%

When the superconducting state is established in the both
subsystems simultaneously and the bare anomalous Green's functions
of conduction electrons are equal to zero the above equations
become more simple:
%
%%%%%%%%%%%%%%%%
\begin{eqnarray}
%%%%%%%%%%%%%%%%
%
G_{\sigma }^{m}(i\omega ) &=&g_{\sigma }(i\omega )G_{\sigma
}^{0}(i\omega ),
\nonumber \\
F_{\sigma \overline{\sigma }}^{m}(i\omega ) &=&f_{\sigma
\overline{\sigma }
}(i\omega )G_{\overline{\sigma }}^{0}(-i\omega ),\label{37} \\
\overline{F}_{\overline{\sigma }\sigma }^{m}(i\omega )
&=&\overline{f}_{ \overline{\sigma }\sigma }(i\omega )G_{\sigma
}^{0}(i\omega ).  \nonumber
%
%%%%%%%%%%%%%%
\end{eqnarray}
%%%%%%%%%%%%%%
%
For the second group of mixed functions in the same conditions we
obtain:
%
%%%%%%%%%%%%%%%%
\begin{eqnarray}
%%%%%%%%%%%%%%%%
%
G_{m\sigma }(i\omega ) &=&G_{\sigma }^{0}(i\omega )g_{\sigma
}(i\omega ),
\nonumber \\
F_{m\sigma \overline{\sigma }}(i\omega ) &=&G_{\sigma }^{0}(i\omega
)f_{\sigma \overline{\sigma }}(i\omega ),\label{38} \\
\overline{F}_{m\overline{\sigma }\sigma }(i\omega )
&=&G_{\overline{\sigma } }^{0}(-i\omega
)\overline{f}_{\overline{\sigma }\sigma }(i\omega ). \nonumber
%
%%%%%%%%%%%%%%
\end{eqnarray}
%%%%%%%%%%%%%%
%
%%%%%%%%%%%%%%%%%%%%%%%%%

\section{Approximations}

%%%%%%%%%%%%%%%%%%%%%%%%%

In previous part of the paper we have formulated the Dyson type
equations for the propagators of the system in general case of
superconducting phase. These equations contain the correlation
functions which take into account charge, spin and pairing
fluctuations and are determined by infinite sums of strong
connected diagrams composed from irreducible Green's functions.
The Dyson type equations for these correlated functions $Z$, $Y$
and $\overline{Y }$ don't exist. Therefore to close the system of
equations and to determine the order parameters of the system
state it is necessary to make some approximations. Our main
approximations are determined by the diagrams shown on the
$Fig.9$.

Our approximations correspond to the summation of ladder diagrams
in vertical direction shown on the $Fig.10$ $a).$ . We neglect the
summation of ladder diagrams in the horizontal direction (see
$Fig.10$ $b)$)

%%%%%%%%%%%%%%%%%%%%%%%%%

\section{Conclusions}

%%%%%%%%%%%%%%%%%%%%%%%%%

The diagrammatic theory has been developed for one-site Anderson model in
which strong correlations of impurity electrons and their hybridization with
conduction electrons is taken into account.

The definition of irreducible Green's functions or Kubo cumulants
is used as a generalized Wick theorem for strongly correlated
subsystem of localized electrons. These irreducible functions
contain all spin, charge and pairing fluctuations. On this base
the linked cluster theorem has been proved to determine the
thermodynamic potential of the system and Dyson type equations
were established for one-particle propagators of the electrons of
both subsystems.The main elements of these equations are the
correlation functions $Z_{\sigma \sigma ^{\prime }}$, $Y_{\sigma
\overline{\sigma }^{\prime }}$ and $\overline{Y}_{
\overline{\sigma }\sigma ^{\prime }}$ which are composed from
strong connected diagrams containing these irreducible Green's
functions.

The normal and superconducting phases are considered. In the last
case we examine the case when only the conduction electron
subsystem has a pairing mechanism of superconductivity and when
the superconductibility is established simultaneously in all the
system as a result of broken symmetry.

%%%%%%%%%%%%%%%%%%%%%%%
\begin{acknowledgments}
The authors have benefited by discussions with Prof. N.M. Plakida and would
like to thank Him. V.A.M. thanks Duisburg -- Essen University for
hospitality and support. This work has been supported in part the Grant of
the Heisenberg -- Landau Program (V.A.M. and P.E.). %%%%%%%%%%%%%%%%%%%%%
\end{acknowledgments}

%%%%%%%%%%%%%%%%%%%%%
%
%%%%%%%%%%%%%%%%%%%%%%%%%%%

%%%%%%%%%%%%%%%%%%%%%

%%%%%%%%%%%%%%


\begin{thebibliography}{99}

%%%%%%%%%%%%%%%%%%%%%%%%%%%
%

\bibitem{Anderson} P.\ W.\ Anderson, Phys.\ Rev.\ \textbf{124}, 41 (1961).

\bibitem{Wolff} P.\ A.\ Wolff, Phys.\ Rev.\ \textbf{124}, 1030 (1961).

\bibitem{Clogston} A.\ M.\ Clogston, Phys.\ Rev.\ \textbf{125}, 439 (1962).

\bibitem{Clogston1} A.\ M.\ Clogston, B.\ T.\ Matthias, M.\ Peter,H.\ J.\
Williams, E.\ Corenzwit and R.\ C.\ Sherwood, Phys.\ Rev.\ \textbf{125},
541(1962).

\bibitem{Clogston2} A.\ M.\ Clogston, Phys.\ Rev.\ \textbf{136}, A1417
(1964).

\bibitem{Klein} A.\ P.\ Klein and A.\ J.\ Heeger, Phys.\ Rev.\ \textbf{144}%
,458 (1966).

\bibitem{Kim} Duk-Joo Kim, Phys.\ Rev.\ \textbf{146}, 455 (1966).

\bibitem{Hamann} D.\ R.\ Hamann, Phys.\ Rev.\ \textbf{158}, 570 (1967).

\bibitem{Bloomfield} P.\ E.\ Bloomfield and D.\ R.\ Hamann, Phys.\ Rev.\
\textbf{164}, 856 (1967).

\bibitem{Theumann} Alba Theumann, Phys.\ Rev.\ \textbf{178}, 978 (1969).

\bibitem{Ueda} K.\ Ueda, Jour.\ Phys.\ Soc.\ Jap.\ \textbf{47}, 811 (1979).

\bibitem{Lacroix} C.\ Lacroix, J.\ Phys.\ F:\ Metal.\ Phys.\ \textbf{11},
2389 (1981).

\bibitem{Wilkins} J.\ W.\ Wilkins, Valence Instabilities, Proceedings of the
International Conference held in Zurick, Switzerland 1982, p.1.

\bibitem{Lacroix1} C.\ Lacroix Valence Instabilities ibid. p. 61.

\bibitem{Fye} R.\ M.\ Fye and J.\ E.\ Hirsch, Phys.\ Rev.\ B.\ \textbf{38},
433 (1988).

\bibitem{Mahan} G.\ D.\ Mahan, Many-Particle Physics, Third Edition, Kluwer
Academic/Plenum Publishers, New York, ch. 6.

\bibitem{Schrieffer} J.\ R.\ Schrieffer and P.\ A.\ Wolff, Phys.\ Rev.\
\textbf{149}, 491 (1966).

\bibitem{Nagaoka} I.\ Nagaoka, Phys.\ Rev.\ \textbf{138}, A1112 (1965).

\bibitem{Falk} D.\ S.\ Falk and M.\ Fowler, Phys.\ Rev.\ \textbf{158}, 567
(1967).

\bibitem{Fowler} M.\ Fowler, Phys.\ Rev.\ \textbf{160}, 463 (1967).

\bibitem{Bogoliubov} N.\ N.\ Bogoliubov and S.\ V.\ Tiablikov, Doklady AN
USSR,\ \textbf{126}, 53 (1959) [in Russian]

\bibitem{Zubarev} D.\ N.\ Zubarev, Usp. Fiz. Nauk,\ \textbf{71}, 71 (1960).

\bibitem{Bonch} V.\ L.\ Bonch-Bruevich and S.\ V.\ Tiablikov, The method of
Quantum Green's Functions of Statistical Physics, Moscow (1961) [in Russian]

\bibitem{Zubarev1} D.\ N.\ Zubarev and Yu.\ A.\ Tserkovnikov, Transaction of
Mathematical Institute V.A. Steklov AN USSR,\ \textbf{175}, 134 (1986) [in
Russian]

\bibitem{Barabanov} A.\ F.\ Barabanov, C.\ A.\ Kikoin and L.\ A.\ Maximov,
Theor.\ Math.\ Phys.\ \textbf{20}, 364 (1974).

\bibitem{Georges} A.\ Georges, G.\ Kotliar, W.\ Krauth and M.\ J.\
Rozenberg, Rev.\ Mod.\ Phys.\ \textbf{8}, 13 (1996).

\bibitem{Kotliar} G.\ Kotliar and D.\ Vollhardt, Physics\ Today\ \textbf{57}%
, 53 (2004).

\bibitem{Hubbard} J.\ Hubbard, Proc.\ Roy.\ Soc.\ \textbf{A276}, 238 (1963).

\bibitem{Hubbard1} J.\ Hubbard, Proc.\ Roy.\ Soc.\ \textbf{A281}, 401 (1964).

\bibitem{Hubbard2} J.\ Hubbard, Proc.\ Roy.\ Soc.\ \textbf{A285}, 542 (1965).

\bibitem{Vladimir} M.\ I.\ Vladimir and V.\ A.\ Moskalenko, Theor.\ Math.\
Phys.\ \textbf{82}, 301 (1990).

\bibitem{Vakaru} S.\ I.\ Vakaru, M.\ I.\ Vladimir and V.\ A.\ Moskalenko,
Theor.\ Math.\ Phys.\ \textbf{85}, 1185 (1990).

\bibitem{Bogoliubov1} N.\ N.\ Bogoliubov and V.\ A.\ Moskalenko, Theor.\
Math.\ Phys.\ \textbf{80}, 10 (1991).

\bibitem{Bogoliubov2} N.\ N.\ Bogoliubov and V.\ A.\ Moskalenko, Theor.\
Math.\ Phys.\ \textbf{92}, 820 (1992).

\bibitem{Moskalenko} V.\ A.\ Moskalenko, P.\ Entel and D.\ F.\ Digor, Phys.\
Rev.\ B\ \textbf{59}, 619 (1999).

%%%%%%%%%%%%%%%%%%%%%
\end{thebibliography}
\end{document}